\documentclass{aa}
\usepackage{epsf}
\usepackage{times}
\newcommand{\md}{\mbox{\rm d}}
\newcommand{\fb}[2]{[#1\,{\sc #2}]}
\newcommand{\al}[2]{#1\,{\sc #2}}
\def\ion#1#2{#1$^{#2}$}
\newcommand{\hb}{\relax\ifmmode{\rm H\beta}\else{\rm H$\beta$}\fi}
\newcommand{\mic}{$\mu$m}
\newcommand{\sctr}[1]{\multicolumn{1}{c}{#1}}
\newcommand{\ctr}[3]{\multicolumn{#1}{c}{#2\hspace*{#3mm}}}

\newcommand{\x}[1]{\hspace*{#1mm}}
\newcommand{\f}{\phantom{0}}
\newcommand{\lrs}{{\it LRS}}
\newcommand{\sws}{{\it SWS}}
\def\nt#1{\vtop{\footnotesize\hsize=\columnwidth\leavevmode#1\hspace*{\fill}}}
\begin{document}
\thesaurus{12(02.01.3; 02.16.1;
09.16.2 NGC~3918; 09.16.2 NGC~6302; 09.16.2 NGC~7027; 13.09.4)}
\title{The collision strength of the [Ne\,{\sc\bf v}] infrared fine-structure
lines\thanks{Based on observations with ISO, an ESA project with instruments
funded by ESA Member States (especially the PI countries: France Germany,
the Netherlands and the United Kingdom) and with the participation of
ISAS and NASA}}
\author{P.A.M.~van~Hoof\inst{1,2}
\and D.A.~Beintema\inst{3}
\and D.A.~Verner\inst{1}
\and G.J.~Ferland\inst{1}}
\offprints{P.A.M. van Hoof at the Toronto address}
\institute{University of Kentucky, Dept.\ of Physics \& Astronomy,
177 CP Building, Lexington, KY 40506--0055, USA
\and Canadian Institute for Theoretical Astrophysics, McLennan Labs, 60 St. George Street,
Toronto, ON M5S~3H8, Canada
\and SRON Laboratory for Space Research, P.O.~Box 800, 9700 AV Groningen,
The Netherlands}
\date{Received date; accepted date}
\maketitle

\begin{abstract}

The calculation of accurate collision strengths for atomic transitions has
been a long standing problem in quantitative spectroscopy.
Most modern calculations are based on the R-matrix method and problems
pertaining to the use of this method have led to a discussion of the accuracy
of these results. More in particular, based on an analysis of the spectra of
NGC~3918 and NGC~6302, Clegg et al.\ (1987) and Oliva et al.\ (1996) have
questioned R-matrix calculations for the infrared \fb{Ne}{v} fine-structure
transitions. Using improved flux measurements for the \fb{Ne}{v} lines, we show
that the conclusion that these collision strengths would be too high, is not
correct. The discrepancies found by Clegg et al.\ (1987) can be explained by
the inaccuracy of the \fb{Ne}{v} 342.6~nm flux they adopted. The discrepancies
found by Oliva et al.\ (1996) can be explained by the inaccuracy of the \lrs\
flux for the \fb{Ne}{v} 14.32~\mic\ line. Based on the data presented in this
paper there is {\em no reason} to assume that there are any problems with the
R-matrix calculations for \ion{Ne}{4+} of Lennon \& Burke (1994). We show that
the data are accurate at the 30~\% level or better. This confirms the validity
of the close coupling method.


\end{abstract}

\vspace*{-24pt}

\section{Introduction}

The calculation of accurate collision strengths for atomic transitions has
been a long standing problem in the field of quantitative spectroscopy. Any
calculation involving atoms in non-LTE conditions requires the knowledge of
vast numbers of collision strengths in order to make these calculations
realistic and accurate. Until recently the computing power was simply not
available to calculate collision strengths in a systematic way. One either had
to resort to simpler and more approximate methods or one had to limit the
calculations to only the most important transitions. This situation has now
changed with the start of the Iron Project (Hummer et al. 1993), which aims to
produce a large database of accurately calculated collision strengths.

The collision strength for an atomic transition depends\\
strong\-ly on the energy
of the colliding electron and shows many resonances (e.g., Aggarwal 1984).
Such resonances occur when the total energy of the target ion and the
colliding electron correspond to an auto-ionizing state. In order to calculate
these resonances accurately, a fine grid of energy points is necessary. This
is a type of problem for which R-matrix methods are very well suited. However,
a source of uncertainty in these calculations is that the energies of most
auto-ionizing states have not been measured in the laboratory and therefore
need to be derived from calculations. It is a well known fact that the
resulting energies are not very accurate and hence the positions of the
resonances are also uncertain. Since the collision strengths are usually
folded with a Maxwellian energy distribution, this is not a major problem for
high temperature (i.e., X-ray) plasmas where the distribution is much broader
than the uncertainty in the position of the resonances. However, for low
temperature (e.g., photo-ionized) plasmas this can lead to problems if a
resonance is present near the threshold energy for the transition. If only the
high-energy tail of the Maxwellian distribution is capable of inducing a
collisional transition, then a small shift in the position of a near-threshold
resonance can have a severe impact on the effective collision strength. This
effect would be even more pronounced if the resonance shifts below the
transition threshold and disappears completely.

\begin{table}
\caption{Comparison of various determinations of the effective collision
strengths for transitions within the ground term of \ion{Ne}{4+}. The values
are for $T_{\rm e}$ = 10\,000~K.}
\label{cstab}
\begin{tabular}{lrrr}
\hline
ref. & $^3\!P_0$ -- $^3\!P_1$ & $^3\!P_0$ -- $^3\!P_2$ & $^3\!P_1$ -- $^3\!P_2$ \\
\hline
Blaha (1969)           & 0.251 & 0.132 & 0.611 \\
Osterbrock (1974)      & 0.23\f& 0.11\f& 0.53\f\\
Aggarwal (1983)        & 1.463 & 1.810 & 5.901 \\
Lennon \& Burke (1991) & 1.401 & 1.766 & 5.725 \\
Lennon \& Burke (1994) & 1.408 & 1.810 & 5.832 \\
\hline
\end{tabular}
\end{table}

The inclusion of resonances in the calculation of collision strengths can lead
to much higher values than were previously published (see Table~2 in Oliva et
al.\ 1996). This is also illustrated in Table~\ref{cstab} where we show
various calculations of the effective collision strength of transitions within
the ground term of \ion{Ne}{4+}. One can see that the R-matrix calculations of
Aggarwal (1983) and Lennon \& Burke (1991, 1994) yield substantially larger
results than the previous calculations. This is caused by the presence of a
large complex of strong resonances at low energies (see Fig.~5 in Aggarwal
1984). This large difference has led to a discussion of the validity the
R-matrix calculations (Clegg et al.\ 1987, C87; Oliva et al.\ 1996, O96). Both
authors tested the calculations by comparing predicted flux ratios with
observations and both concluded that the R-matrix calculations yielded results
that are too high. Nebulae offer powerful tests of atomic physics, and have
revealed incomplete treatment in the past (P\'equignot et al.\ 1978,
Harrington et al. 1980). However, both C87 and O96 base their conclusions on
only one nebula and both include \lrs\ data in their analysis. Since the
accuracy of \lrs\ data is limited to approximately 30~\% (Pottasch et al.
1986), a re-analysis based on more accurate \sws\ data is warranted. In this
paper we will present a test of the R-matrix calculations for \ion{Ne}{4+} by
applying them to observational data of NGC~3918, NGC~6302 (the nebulae studied
by C87 and O96, respectively) and NGC~7027. This will yield a larger and more
accurate sample for the discussion.

\section{The observational data and analysis}

\begin{table}
\caption{Log of the \sws\ observations. The AOT's are described in de Graauw
et al. (1996), SWS01--4 stands for SWS01 speed~4.}
\label{tabsws}
\scriptsize
\begin{tabular}{lrrlcc}
\hline
Source   & \sctr{Date} & \sctr{Rev.} & \sctr{AOT} & $F(14.32)$ & $F(24.32)$ \\
         &             &             & & \ctr{2}{10$^{-14}$~W\,m$^{-2}$}{0} \\
\hline
NGC~7027 & 11 Dec 1995 &  24 & SWS01--4 & 160. & 47.0 \\
NGC~7027 & 19 Dec 1995 &  32 & SWS02    & 136. & 40.9 \\
NGC~7027 &  6 Nov 1996 & 356 & SWS02    & 141. & 47.1 \\
\\		      
NGC~6302 & 19 Feb 1996 &  94 & SWS01--4 & 65.2 & 29.7 \\
NGC~6302 & 20 Feb 1997 & 462 & SWS06    & 63.4 & 30.8 \\
\hline
\end{tabular}
\end{table}

Several \sws\ observations were obtained for the objects studied in this
paper. A log of the observations is shown in Table~\ref{tabsws}. The
instrument is described in de Graauw et al. (1996). The observations used
three different templates: SWS01 -- a spectral scan from 2.4~\mic\ to 45~\mic,
SWS02 -- a set of grating scans of individual lines, and SWS06 -- a high
resolution grating scan. All \sws\ spectra of NGC~7027 were obtained during
calibration time. The complete SWS06 spectrum of NGC~6302 is published in
Beintema \& Pottasch (1999). An \sws\ spectrum of NGC~3918 was also
obtained, but not used. Due to inaccurate pointing the source was partially
outside the aperture.

\begin{table}
\caption{The values for the line fluxes (corrected for reddening) and the
extinction adopted in this work. Entries in italics are assumed. The rows
labeled O96 and C87 give the values adopted by Oliva et al. (1996) and Clegg
et al. (1987).}
\label{tab}
\begin{tabular}{lrrrr}
\hline
source   &\sctr{$F(342.6)$}&\sctr{$F(14.32)$}&\sctr{$F(24.32)$}&\sctr{$c(\hb)$}\\
         &         \ctr{3}{10$^{-14}$~W\,m$^{-2}$}{0}         &   \sctr{dex}  \\
\hline
NGC~7027 & 204.$^a$\f & 153.\f\f  & 46.9\f    & 1.37$\pm$0.05$^a$ \\
\\
NGC~6302 &  70.$^b$\f & 67.2\f    & 31.4\f    & 1.23$\pm$0.10$^c$ \\
O96      &  86.$^c$\f & 45.7$^d$  & 32.3$^d$  & 1.23$\pm$0.10$^c$ \\
\\
NGC~3918 &  7.4$^e$\rlap{:} & 12.$^f$\f & \it 8.1\f & 0.43$\pm$0.05$^g$ \\
C87      &  20.$^g$\f & 12.$^f$\f &    ---\f  & 0.43$\pm$0.05$^g$ \\
\hline
\end{tabular}
\par
\nt{\scriptsize%
$^a$~Middlemass (1990)\ \ 
$^b$~Beintema \& Pottasch (1999)\ \
$^c$~Oliva et al. (1996)\ \ 
$^d$~Rowlands et al. (1994)\ \ 
$^e$~Aller \& Faulkner (1964), corrected for blend\ \
$^f$~Pottasch et al. (1986)\ \ 
$^g$~Clegg et al. (1987)}
\end{table}

The line fluxes were measured in spectra reduced with the \sws\
interactive-analysis software. The line fluxes were virtually identical to
those derived from the standard {\it ISO} auto-analysis products. The various
\sws\ measurements were subsequently averaged. Table~\ref{tab} shows the
dereddened line fluxes we have adopted for our study. The values adopted by
C87 and O96 are also shown for comparison. The extinction corrected \fb{Ne}{v}
342.6~nm fluxes were taken from the original publications. Both for NGC~7027
and NGC~6302 the dereddening is complicated by the fact that the extinction
varies over the nebula. The correction for the blend in NGC~3918 is discussed
below. The infrared line fluxes were dereddened using the law from Mathis
(1990).

\begin{figure*}
\begin{center}
\mbox{\epsfxsize=0.60\textwidth\epsfbox[22 344 557 721]{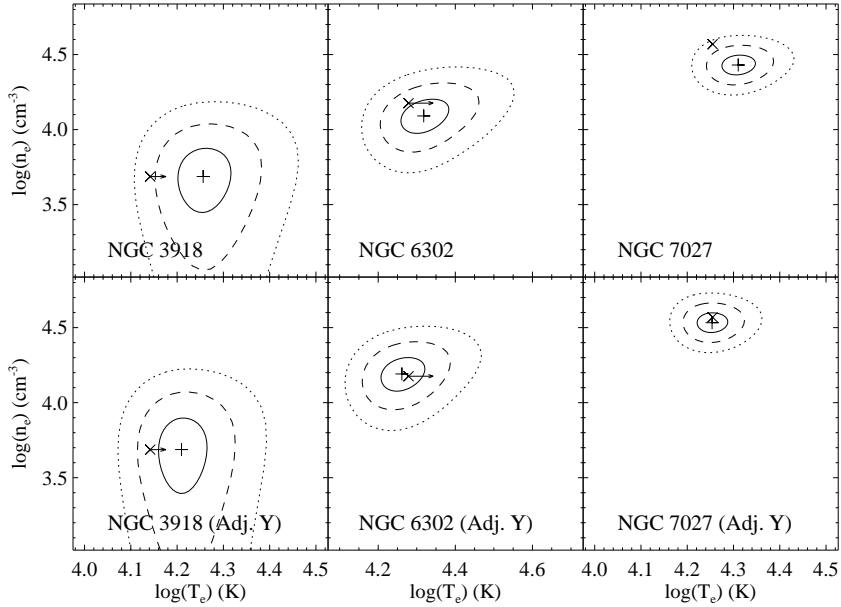}}
\end{center}
\caption{In these graphs the electron temperature and density derived from the
\fb{Ne}{v} lines is marked with a $+$. The solid, dashed and dotted lines are
the 1~$\sigma$, 2~$\sigma$ and 3~$\sigma$ contours respectively. The expected
value of the electron temperature and density, based on either
photo-ionization modeling or other line-ratio diagnostics, is indicated with
an $\times$. An arrow indicates the estimated increase of the electron
temperature if temperature stratification is taken into account. The top row
was obtained using the effective collision strengths $\Upsilon$ given by
Lennon \& Burke (1994), the bottom row was obtained by lowering the
$\Upsilon$'s for transitions within the ground term by 30~\%.}
\label{nev:fig}
\end{figure*}

To calculate the diagnostic diagrams we used the effective collision strengths
given in Lennon \& Burke (1994) and adopted the transition probabilities from
Baluja (1985). We used a 5-level atom to calculate the relative level
populations. The results of our analysis are shown in Table~\ref{tab:res} and
Fig.~\ref{nev:fig}. The line flux ratios are defined as: $R_1 =
I(14.32)/I(342.6)$ and $R_2 = I(14.32)/I(24.32)$. To determine the
uncertainties in the line ratios we included contributions from the absolute
calibration accuracy of the UV and IR data, the internal calibration accuracy
of the SWS data and the uncertainty in the extinction correction. We will now
discuss the results for each nebula separately.

\begin{table}
\caption{This table first shows the observed line flux ratios, $R_1$ and
$R_2$. Next it shows the derived values for the electron temperature and
density and finally the expected values for these quantities. Entries in
italics are assumed.}
\label{tab:res}
\begin{tabular}{lrrr}
\hline
                            & NGC~3918 & NGC~6302 & NGC~7027 \\
\hline
$R_1$                       &    1.65$\pm$0.74 &    0.96$\pm$0.42 &    0.75$\pm$0.17 \\
$R_2$                       &\it 1.50          &    2.14$\pm$0.29 &    3.26$\pm$0.29 \\
\\
$T_{\rm e}$/K               & 18\,000$\pm$2400 & 20\,800$\pm$3000 & 20\,400$\pm$1700 \\
$n_{\rm e}$/cm$^{-3}$       &   \it 4\,900     & 12\,300$\pm$3300 & 26\,900$\pm$4100 \\
\\
$T_{\rm e}$/K (exp.)        & 13\,900          & 19\,000          & 18\,000          \\
$n_{\rm e}$/cm$^{-3}$ (exp.)&  4\,900          & 15\,000          & 37\,100          \\
\hline
\end{tabular}
\end{table}

\subsection{NGC~7027}
\label{neb:i}

We have included this nebula in our sample because it is bright and well
studied. This assures that accurate values for the fluxes needed in our
analysis are available. Additionally, we have a detailed photo-ionization
model (Beintema et al.\ 1996) which allows us to derive good estimates for the
expected electron temperature and density. For a high excitation nebula like
NGC 7027, significant temperature stratification of the plasma can be
expected. We have illustrated this effect in Table~\ref{strat} where we show
the average electron temperature in various line forming regions. One can see
that the electron temperature rises almost monotonically with the ionization
potential $\chi_{\rm ion}$. It is important to note that the temperature in
the \ion{Ne}{4+} region is substantially higher than the temperature in any
other line forming region observed in the spectrum. The expected values for
the electron temperature and density are based on our photo-ionization model.
The analysis of the \fb{Ne}{v} lines gives a result which deviates less than
3~$\sigma$ from these values.

\begin{table}
\caption{The temperature stratification in NGC~7027, derived from our
photo-ionization model (Beintema et al.\ 1996). The temperature is
defined as $\int n_{\rm e} n_{\rm ion} T_{\rm e} \md V /
\int n_{\rm e} n_{\rm ion} \md V$.}
\label{strat}
\begin{tabular}{lrr@{\x{8}}lrr}
\hline
ion & \sctr{$\chi_{\rm ion}$/eV} & \ctr{1}{$T_{\rm e}$/K}{6} & ion & \sctr{$\chi_{\rm ion}$/eV} & \sctr{$T_{\rm e}$/K} \\
\hline
\ion{O}{0}   & 13.62 & 10\,450 & \ion{O}{2+}  & 54.94 &  13\,520 \\
\ion{N}{+}   & 29.60 & 12\,140 & \ion{Ar}{3+} & 59.81 &  14\,670 \\
\ion{Ar}{2+} & 40.74 & 12\,990 & \ion{Ar}{4+} & 75.02 &  16\,180 \\
\ion{S}{2+}  & 34.83 & 13\,100 & \ion{Ne}{3+} & 97.12 &  16\,350 \\
\ion{C}{2+}  & 47.89 & 13\,440 & \ion{Ne}{4+} &126.2\f&  17\,990 \\
\hline
\end{tabular}
\end{table}

\subsection{NGC~6302}

This is the nebula studied by O96.
The \fb{Ne}{v} temperature and density for this nebula were already derived
from the SWS06 spectrum by Pottasch \& Beintema (1999). In our analysis we
will include the SWS01--4 spectrum as well. The expected values
for the electron temperature and density are those of O96. Our analysis gives a
result which deviates slightly more than 1~$\sigma$ from these values. The
preferred temperature of O96 is based on various determinations using ions
with lower ionization potentials than \ion{Ne}{4+}. In view of the discussion
in the previous section concerning temperature stratification, this estimate
is probably too low. Especially in view of the temperatures derived by O96
from rather low excitation line ratios like \fb{S}{iii}, \fb{Ar}{iii} and
\fb{O}{iii} which range between 18\,100~K and 19\,400~K, the \fb{Ne}{v}
temperature may be expected to be considerably higher than 19\,000~K. A
temperature of 22\,000~K is more realistic (see the discussion in Pottasch \&
Beintema 1999). This value for the temperature is indicated by an arrow in
Fig.~\ref{nev:fig}. After this correction the discrepancy is slightly less
than 1~$\sigma$.

\subsection{NGC~3918}
\label{neb:iii}

This is the nebula studied by C87.
The intensity for the \fb{Ne}{v} 342.6~nm line is in doubt. C87 quote
$I(342.6) = 80$, but it is not clear how this value was obtained. We decided
to use the value quoted in Aller \& Faulkner (\cite{af64}) instead. From the
discussion in that article it is not clear whether the data were corrected for
interstellar extinction. The intensities they quote for other strong blue
emission lines compare well with the dereddened intensities given by C87 and
we therefore assume that the Aller \& Faulkner (\cite{af64}) data are
corrected for interstellar extinction. They give $I(342.6+342.9+344.4) = 60$.
The correction for the blend with the \al{O}{iii} 342.9~nm and 344.4~nm Bowen
resonance-fluorescence lines is easy, since the \al{O}{iii} 313.3~nm line has
been measured by C87. The \al{O}{iii} 313.3~nm, 342.9~nm and 344.4~nm lines
all originate from the same upper level ($2s^2\,2p\,3d$ $^3\!P^\circ_2$) and the
intensity ratio of the lines is simply given by the ratio of the transition
probabilities times the photon energy. Using Opacity Project data (Luo et al.\
1989) one finds $I(313.3)$~: $I(342.9)$~: $I(344.4)$~= 10.94~: 1.00~: 2.94.
C87 gives $I(313.3) = 85$. Hence $I(342.9+344.4) = 30.6$
and $I(342.6) = 29.4$. This result is substantially lower than the value used
by C87. We were not able to correct the \sws\ spectrum accurately for aperture
effects and therefore preferred to use the \lrs\ flux for the \fb{Ne}{v}
14.32~\mic\ line. To complete the data set, we assumed a value for the
24.32~\mic\ flux such that the resulting density agreed with the expected
value. None of the flux values we adopted for this nebula can be considered
accurate and re-measurement is warranted. The expected values for the electron
temperature and density were determined by averaging the data in Table~12 of
C87. For the temperature we only used the values derived from the \fb{Ar}{v},
\fb{Ne}{iv} and \fb{Ne}{v} line ratios, for the density we used all values
except those derived from \fb{Mg}{i}, \fb{N}{iv} and \fb{O}{iv} lines. One can
see that there is a slightly more than 2~$\sigma$ discrepancy for the electron
temperature. Again the expected value for the electron temperature may be
underestimated due to temperature stratification. We think 15\,000~K is in all
probability a more realistic, though still conservative, estimate. This value
for the temperature is indicated by an arrow in Fig.~\ref{nev:fig}. After this
correction there is a 1.5~$\sigma$ discrepancy.

\section{Discussion}

A major advance in atomic theory in the past decade has been close coupling
calculations that include resonances. These new calculations (carried out with
an R-matrix code) can raise the collision strength by an order of magnitude or
more compared to older calculations. These large differences have led to a
discussion of the validity of these calculations. Nebulae offer powerful
laboratories for verifying atomic processes, and two studies (C87 and O96)
used this approach to test the R-matrix calculations for \ion{Ne}{4+}. They
found that spectra of planetary nebulae did not agree with the R-matrix
results. This casted doubt on the validity of the close coupling calculations.

In this paper we have redone the analysis carried out by C87 and O96 using
newer, more accurate data. We also included the well studied nebula NGC~7027
in the analysis to obtain a larger sample. We found that the expected values
of the electron temperature and density all were within 3~$\sigma$ of our
results. Hence there is no proof for significant problems with the R-matrix
calculations. On closer inspection one sees that the largest discrepancy is
for NGC~7027, the best studied nebula. Also the electron temperature derived
from our analysis is systematically higher than the expectation value for all
nebulae. This could point to inaccuracies in the collision strengths. To check
this point further, we have re-analyzed our sample using effective collision
strengths which were lowered by 30~\% for transitions within the $^3\!P$
ground term. The results are shown in the bottom row of Fig.~\ref{nev:fig}.
They are in good agreement with the expected values. Hence the R-matrix
calculations for \ion{Ne}{4+} could be off by 30~\%, but certainly not more.
We point out that this is significantly less than the factor $\sim$2.7 suggested
by O96. An alternative explanation could be that the \fb{Ne}{v} 342.6~nm fluxes
are systematically overestimated due to a problem with the extinction curve,
or a combination of the two effects.

We reach the following main conclusions:

\begin{enumerate}
\item
The discrepancies found by C87 can be explained by the inaccuracy of the
\fb{Ne}{v} 342.6~nm flux they adopted. The discrepancies found by O96 mainly
stem from the inaccuracy of the \lrs\ measurement of the \fb{Ne}{v}
14.32~\mic\ line.
\item
Based on the data presented in this paper there is {\em no reason} to assume
that there are any problems with the collision strengths for \ion{Ne}{4+}
calculated by Lennon \& Burke (1994). Our analysis has shown that the data are
accurate at the 30~\% level or better. This confirms the validity of close
coupling calculations.
\end{enumerate}

\acknowledgements{In this paper data from the Atomic Line List v2.02
(\verb$http://www.pa.uky.edu/~peter/atomic$) were used.
We thank the NSF and NASA for support through grants AST 96-17083 and
GSFC--123.}

\end{document}